\documentclass[]{spie}  

\usepackage{amsmath,amsfonts,amssymb}
\usepackage{graphicx}
\usepackage[colorlinks=true, allcolors=blue]{hyperref}

\title{Thermal Testing for Cryogenic CMB Instrument\\ Optical Design}

\author[a]{D.C. Goldfinger}
\author[b]{P.A.R. Ade}
\author[c]{Z. Ahmed}
\author[d]{M. Amiri}
\author[a]{D. Barkats}
\author[e]{R. Basu Thakur}
\author[c, f]{D. Beck}
\author[g]{C.A. Bischoff}
\author[e, h]{J.J. Bock}
\author[i]{V. Buza}
\author[j]{J. Cheshire}
\author[k]{J. Connors}
\author[a]{J. Cornelison}
\author[l]{M. Crumrine}
\author[f, c]{A.J. Cukierman}
\author[k]{E.V. Denison}
\author[a]{M.I. Dierickx}
\author[m]{L. Duband}
\author[a]{M. Eiben}
\author[d]{S. Fatigoni}
\author[n, o]{J.P. Filippini}
\author[g]{C. Giannakopoulos}
\author[f]{N. Goeckner-Wald}
\author[f]{J. Grayson}
\author[a]{P.K. Grimes}
\author[l]{G. Hall}
\author[f]{G. Halal}
\author[d]{M. Halpern}
\author[g]{E. Hand}
\author[a]{S.A. Harrison}
\author[c]{S. Henderson}
\author[e, h]{S.R. Hildebrandt}
\author[k]{G.C. Hilton}
\author[k]{J. Hubmayr}
\author[e]{H. Hui}
\author[f, c, k]{K.D. Irwin}
\author[f, e]{J. Kang}
\author[a, i]{K.S. Karkare}
\author[e]{S. Kefeli}
\author[a, p]{J.M. Kovac}
\author[f, c]{C.L. Kuo}
\author[l]{K. Lau}
\author[i]{E.M. Leitch}
\author[n]{A. Lennox}
\author[f]{T. Liu}
\author[h]{K.G. Megerian}
\author[e]{L. Minutolo}
\author[e]{L. Moncelsi}
\author[f]{Y. Nakato}
\author[q]{T. Namikawa}
\author[h]{H.T. Nguyen}
\author[e, h]{R. O'Brient}
\author[g]{S. Palladino}
\author[a]{M.A. Petroff}
\author[m]{T. Prouve}
\author[l, j]{C. Pryke}
\author[a ,r]{B. Racine}
\author[k]{C.D. Reintsema}
\author[f]{M. Salatino}
\author[e]{A. Schillaci}
\author[a]{B.L. Schmitt}
\author[j]{B. Singari}
\author[e]{A. Soliman}
\author[a]{A.G. Smith}
\author[a, p]{T. St. Germaine}
\author[e]{B. Steinbach}
\author[b]{R.V. Sudiwala}
\author[f, c]{K.L. Thompson}
\author[a]{C. Tsai}
\author[b]{C. Tucker}
\author[h]{A.D. Turner}
\author[g, n]{C. Umilt\`{a}}
\author[a]{C. Verg\`{e}s}
\author[s, i]{A.G. Vieregg}
\author[e]{A. Wandui}
\author[h]{A.C. Weber}
\author[d]{D.V. Wiebe}
\author[l]{J. Willmert}
\author[c]{W.L.K. Wu}
\author[f]{H. Yang}
\author[f, c]{K.W. Yoon}
\author[f, c]{E. Young}
\author[f]{C. Yu}
\author[a]{L. Zeng}
\author[e]{C. Zhang}
\author[e]{S. Zhang}

\affil[a]{Center for Astrophysics, Harvard \& Smithsonian, Cambridge, Massachusetts 02138, U.S.A}
\affil[b]{School of Physics and Astronomy, Cardiff University, Cardiff, CF24 3AA, United Kingdom}
\affil[c]{Kavli Institute for Particle Astrophysics and Cosmology, SLAC National Accelerator Laboratory, Menlo Park, California 94025, USA}
\affil[d]{Department of Physics and Astronomy, University of British Columbia, Vancouver, British Columbia V6T 1Z1, Canada}
\affil[e]{Department of Physics, California Institute of Technology, Pasadena, California 91125, USA}
\affil[f]{Department of Physics, Stanford University, Stanford, California 94305, USA}
\affil[g]{Department of Physics, University of Cincinnati, Cincinnati, Ohio 45221, USA}
\affil[h]{Jet Propulsion Laboratory, Pasadena, California 91109, USA}
\affil[i]{Kavli Institute for Cosmological Physics, University of Chicago, Chicago, Illinois 60637, USA}
\affil[j]{Minnesota Institute for Astrophysics, University of Minnesota, Minneapolis, Minnesota, 55455, USA}
\affil[k]{National Institute of Standards and Technology, Boulder, Colorado 80305, USA}
\affil[l]{School of Physics and Astronomy, University of Minnesota, Minneapolis, Minnesota 55455, USA}
\affil[m]{Service des Basses Temp\'{e}ratures, Commissariat \`{a} l'Energie Atomique, 38054 Grenoble, France}
\affil[n]{Department of Physics, University of Illinois at Urbana-Champaign, Urbana, Illinois 61801, USA}
\affil[o]{Department of Astronomy, University of Illinois at Urbana-Champaign, Urbana, Illinois 61801, USA}
\affil[p]{Department of Physics, Harvard University, Cambridge, Massachusetts 02138, USA}
\affil[q]{Kavli Institute for the Physics and Mathematics of the Universe (WPI), UTIAS, The University of Tokyo, Kashiwa, Chiba 277-8583, Japan}
\affil[r]{Aix-Marseille Universit\'{e}, CNRS/IN2P3, CPPM, 13288 Marseille, France}
\affil[s]{Department of Physics, Enrico Fermi Institute, University of Chicago, Chicago, Illinois 60637, USA}

\authorinfo{Further author information: (Send correspondence to D.C.G..)\\D.C.G.: E-mail: david.goldfinger@cfa.harvard.edu}

\begin{document} 
\maketitle

\begin{abstract}
Observations of the Cosmic Microwave Background rely on cryogenic instrumentation with cold detectors, readout, and optics providing the low noise performance and instrumental stability required to make more sensitive measurements.  It is therefore critical to optimize all aspects of the cryogenic design to achieve the necessary performance, with low temperature components and acceptable system cooling requirements.  In particular, we will focus on our use of thermal filters and cold optics, which reduce the thermal load passed along to the cryogenic stages.  To test their performance, we have made a series of in situ measurements while integrating the third receiver for the BICEP Array telescope.  In addition to characterizing the behavior of this receiver, these measurements continue to refine the models that are being used to inform design choices being made for future instruments.
\end{abstract}


\section{Background}
\label{sec:intro} 
Measurements of the Cosmic Microwave Background (CMB) play a critical role in our study of the early universe.  Models of cosmic inflation predict that primordial gravitational waves will have imprinted a faint signal in the B-mode spectrum of the CMB.  To measure this polarization signal, we require instrumentation that can provide high throughput (through large optical field of view and detector count) as well detectors with high sensitivity to the CMB signal.  By using cryogenic instrumentation, we are able to achieve low noise performance and stability of the instrument response.  In particular, the cryogenic optics design provides the dual benefit of reducing the thermal loading, allowing the colder components of the instrument to be more effectively cooled, as well as reducing the portion of that loading that is in the detector band, which increases the detector sensitivity to the CMB signal.

\begin{figure} [t]
   \begin{center}
   \begin{tabular}{c} 
   \includegraphics[width=0.9\linewidth]{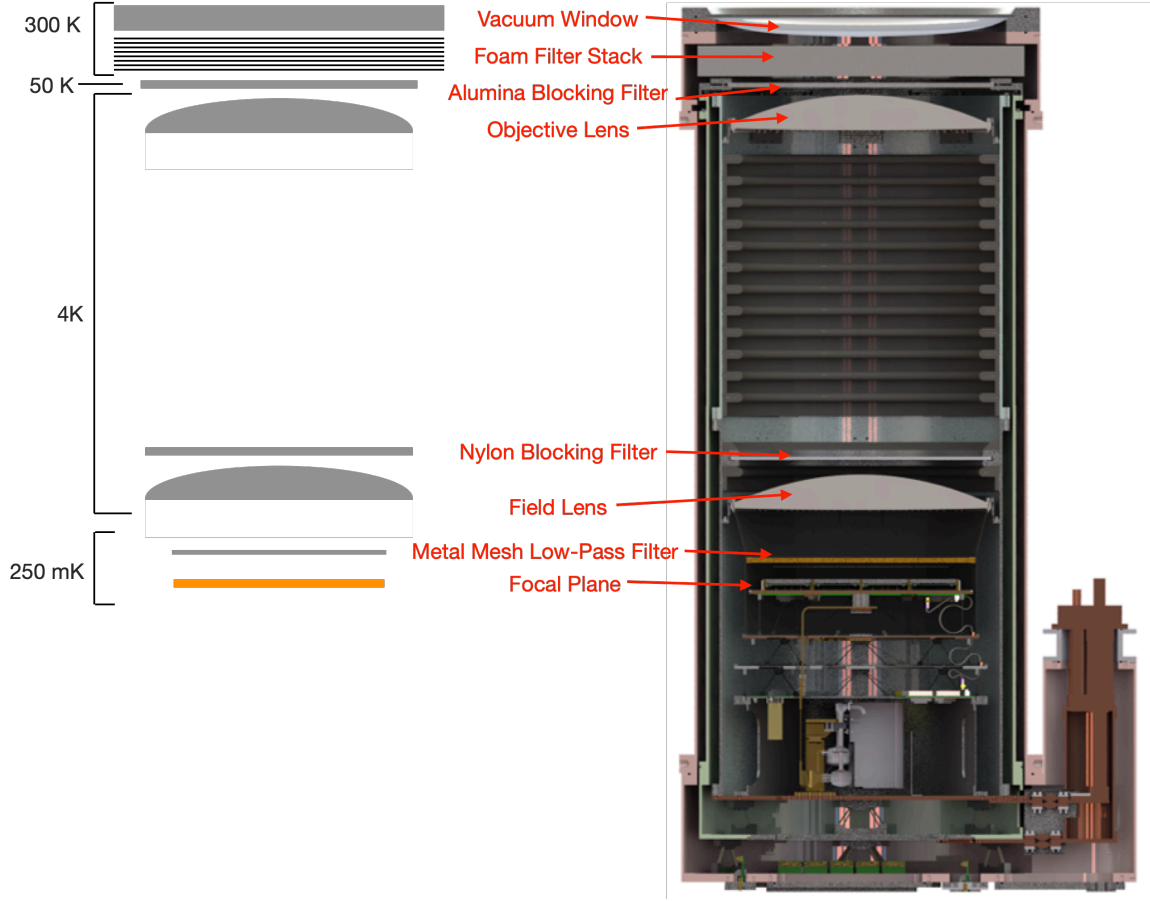}
   \end{tabular}
   \end{center}
   \caption[example] 
   { \label{fig:design} 
Left: Schematic cross section of the BICEP Array cryostat, including only the optical components included in the thermal model.  Right: Cross section image of the BICEP Array solid model.  The foam filter stack is supported from the room temperature stage, along with the vacuum window.  The alumina blocking filter is held on the top of the 50K shell and the lenses and nylon filter are support in the 4K shell.  The metal mesh filter is supported by the sub-Kelvin assembly above the focal plane.}
   \end{figure}

In order to achieve these sensitive measurements, the BICEP Array (BA) telescope consists of four receivers that will operate at the South Pole, each with their own cryogenic system \cite{Crumrine_2018}.  The first of these receivers has been deployed, demonstrating the necessary cryogenic performance for science operations \cite{Moncelsi_2020}.  The cryostat design (Fig. \ref{fig:design}) consists of concentric cylinders at nominal temperatures of 50K and 4K, inside of a vacuum jacket, cooled by a Cryomech PT415 pulse tube.  A He-3 sorption fridge supported from the 4K stage provides the cooling power for stages at 2K, 350 mK, and 250 mK, with the instrument’s superconducting detectors on the coldest stage.  Conductive loads between each of these stages are minimized through the use of low thermal conductivity materials in the support structure (carbon fiber, G-10, and thin Titanium) as well as in the readout wiring (manganin and NbTi), while the quantity of readout wiring is reduced through the use of multiplexed readout \cite{Crumrine_2018}.  Convective loads are minimized through vacuum inside the cryostat and low radiative loads between the walls of the thermal stages are achieved through multilayer insulation \cite{Crumrine_thesis}.

Our design calculations indicate that the largest component to the thermal loading in the cryostat arises from thermal radiation that passes through the optics \cite{Crumrine_2018}.  The filtering materials chosen in this design serve as low pass filters, so the higher frequency infrared radiation is absorbed while the lower frequency CMB signal is allowed to pass through.  Each optical component will then radiate its own blackbody spectrum, with a lower total load commensurate with its lower source temperature.  The BA optics design (Fig. \ref{fig:design}) starts with an HDPE window on the vacuum jacket, with a stack of 12 $1/8$ inch thick Zotefoams Plastazote HD-30 filters.  Due to the low thermal conductivity of the foam, the heat flow into the individual foam layers is dominated by radiation from the adjacent layers.  As the stack comes into radiative equilibrium, the temperature decreases through the stack so that the net radiation passed on to the 50K stage is reduced.  An alumina filter provides another stage of thermal filtering on the 50K stage, above the HDPE lenses on the 4K stage.  Also on the 4K stage (either between or below the lenses, depending on the particular receiver's optical design) a nylon filter provides another stage of thermal filtering before reaching the final filtering on the subKelvin stages.  To fully understand the effectiveness of this filtering scheme, we have developed a thermal model for the optics, which has been refined and tested through direct measurements in BICEP Array receivers.

\section{Thermal Modeling}
\label{sec:model} 
To model the thermal load through the optics, we consider the radiative balance equation for each element (depicted in Fig. \ref{fig:balance}):

   \begin{figure} [t]
   \begin{center}
   \begin{tabular}{c} 
   \includegraphics[height=5cm]{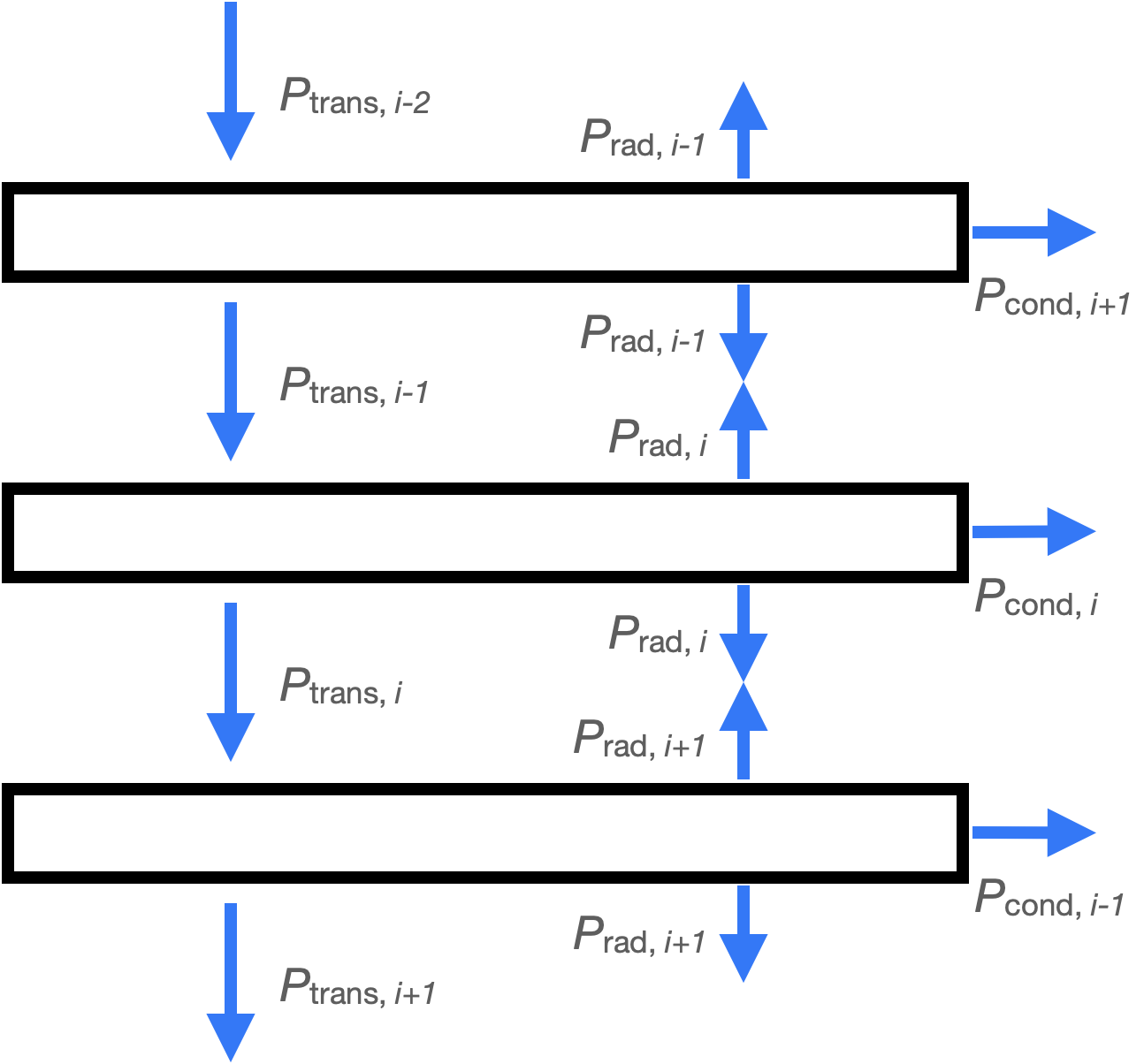}
   \end{tabular}
   \end{center}
   \caption[example] 
   { \label{fig:balance} 
Diagram representing multiple components of the thermal model and the heat flows that pass between them.}
   \end{figure}

\begin{equation}
P_{\text{trans}, i-1} + P_{\text{rad}, i-1} +  P_{\text{rad}, i+1} - P_{\text{trans}, i} = P_{\text{cond}, i} + 2 \times P_{\text{rad}, i}
\end{equation}

$P_{\text{trans}, i}$ indicates the power transmitted through the $i$th element of the optics, $P_{\text{rad}, i}$ is the power radiated by that stage, and $P_{\text{cond}, i} $ is the heat conducted out through the support at the edge of the element.  In the case of all elements other than the stack of foam filters, we can say that $P_{\text{rad}, i-1} \gg  P_{\text{rad}, i+1}$ since the temperature of the $i-1$ layer is significantly colder than the $i+1$ layer.  Therefore our model follows:

\begin{equation}
P_{\text{trans}, i-1} + P_{\text{rad}, i-1} - P_{\text{trans}, i} = P_{\text{cond}, i} + 2 \times P_{\text{rad}, i}
\end{equation}

\begin{figure} [t]
   \begin{center}
   \begin{tabular}{c} 
   \includegraphics[height=4cm]{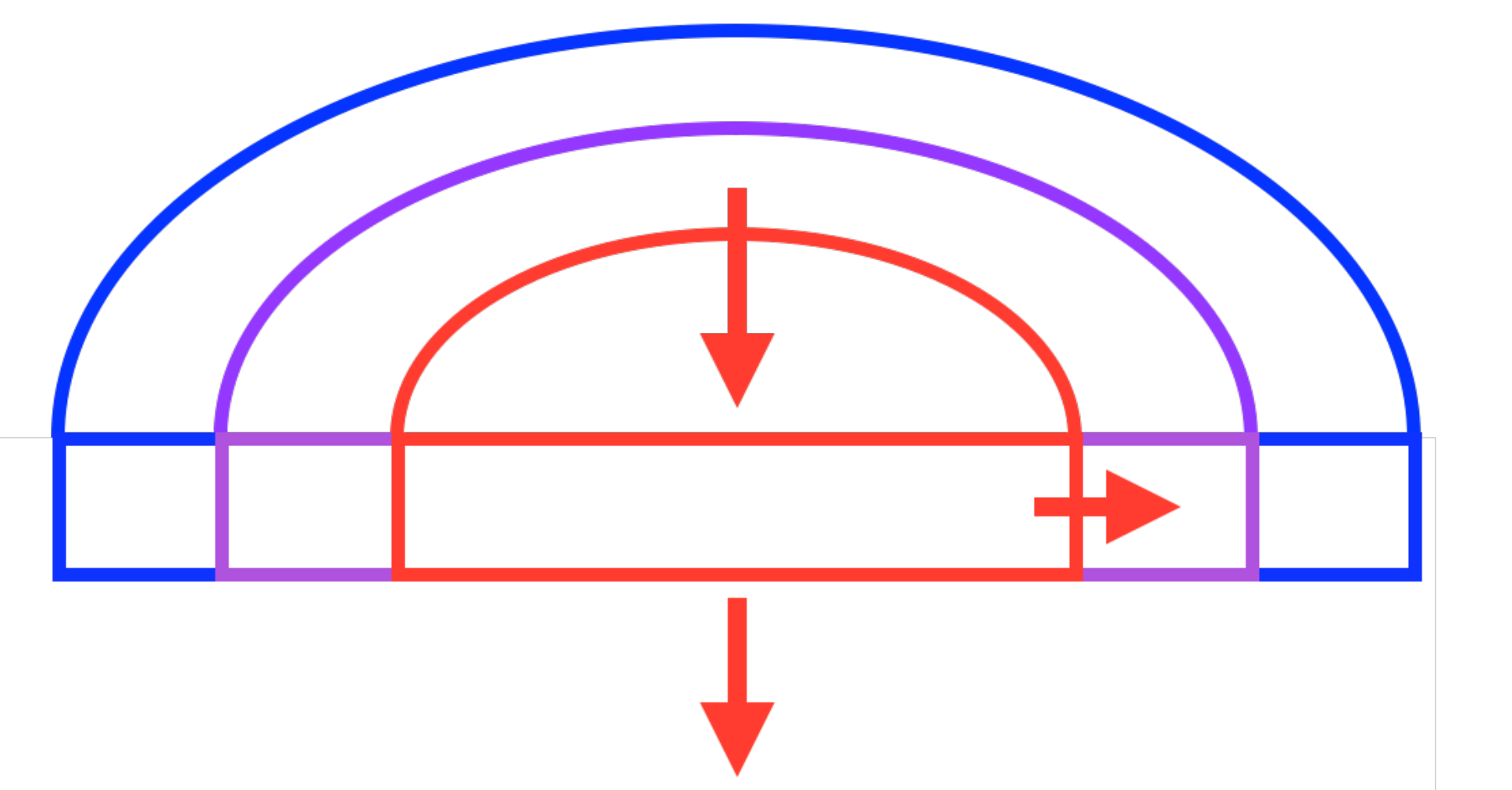}
   \end{tabular}
   \end{center}
   \caption[example] 
   { \label{fig:shells} 
Cross sectional diagram representing an optical component in the thermal model, where thermal balance is calculated between radiation absorbed within a certain radius, power conducted to the edge, and radiation emitted.}
   \end{figure}

At each stage, we model the (cylindrically symmetric) elements as discrete rings with known radius and thickness and a variable temperature (Fig. \ref{fig:shells}).  We can then calculate the energy balance for each ring using the total power that was absorbed and emitted in the portion of the disk interior to a particular ring, and the conducted power through the outer edge of the ring.  The absorbed power is taken from the incident loads, with the fractions absorbed and transmitted calculated from the transmission spectrum of incident radiation and frequency dependent transmission data for the particular material.  To calculate that average optical depth for the transmission coefficient, we use the diffuse approximation to describe the transmission between layers, as they are very close together. The model then calculates the temperature gradient that is necessary to achieve thermal equilibrium, as the conducted power is calculated from the material conductance and modeled temperature difference, with the edge temperatures constrained by model inputs (taken from direct measurement in the cryostat) and the radiated power comes from the component transmittance and temperatures.  Once a temperature gradient that minimizes the residuals in the energy balance for a given layer is found, an effective temperature is calculated to provide the radiated power for the next layer in the stack.

\begin{figure} [t]
   \begin{center}
   \begin{tabular}{c} 
   \includegraphics[height=7cm]{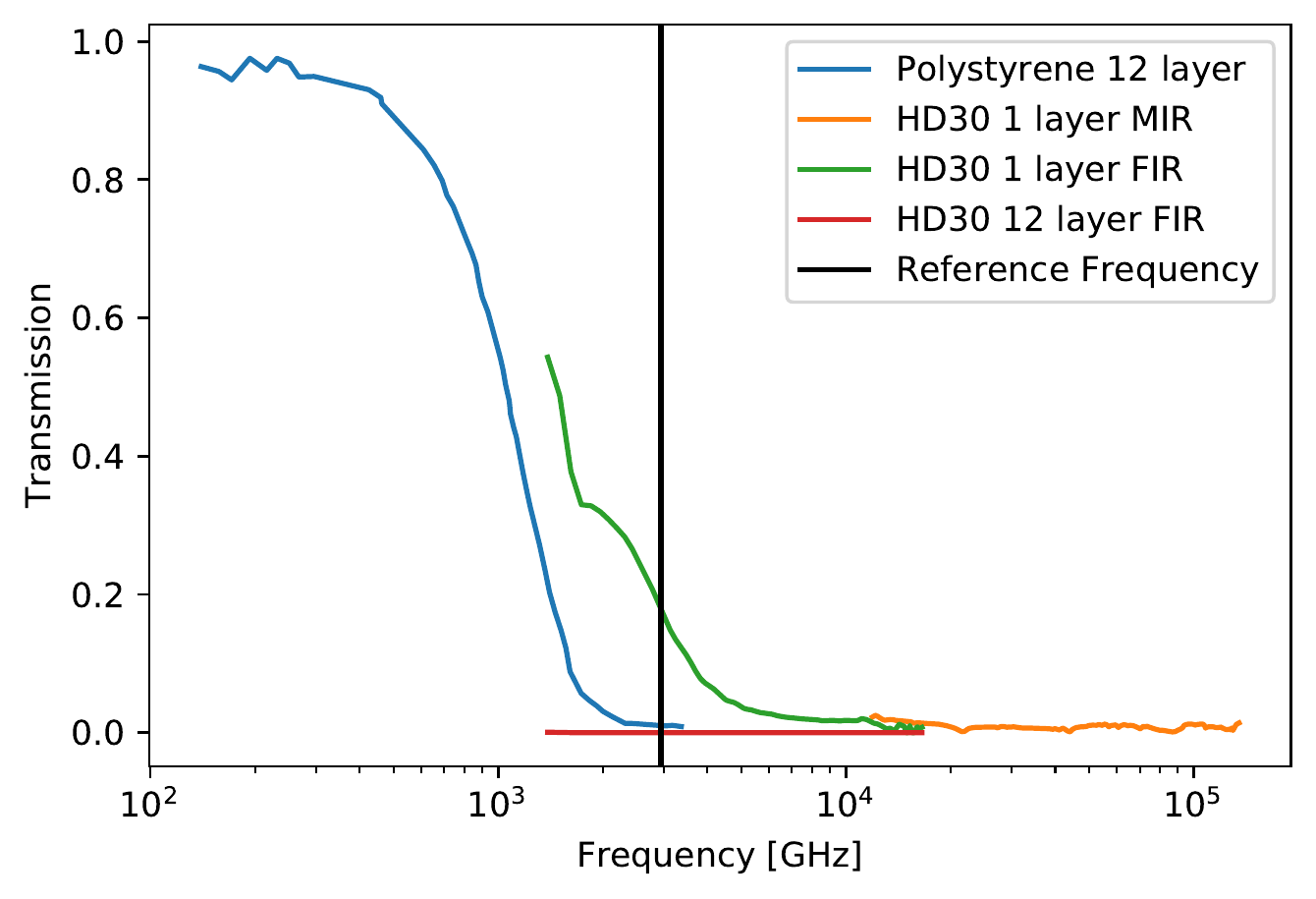}
   \end{tabular}
   \end{center}
   \caption[example] 
   { \label{fig:hd30fts} 
Transmission of a single HD-30 foam filter, in an FTS setup, taken at room temperature \cite{Wandui_thesis}.  The reference frequency of 3 THz indicates the frequency below which 300 K radiation transmitted through the stack can be considered small in comparison to the infrared radiation emitted by the bottom layer of the foam filter stack.  Comparison data for Polystyrene is taken from the Choi et al. reference \cite{Choi_2013}.}
   \end{figure}

For the multilayer foam filter stack, the load is modeled separately via the method discussed in Choi et al.\cite{Choi_2013}.  Because the foam material has a low thermal conductivity, the conduction out the sides of the filters can be neglected, allowing the filter temperatures to be modeled solely through radiative equilibrium.  However, the warm stages are sufficiently close together in temperature that $P_{\text{rad}, i+1}$ is significant for the load on each filter, so the model described above does not directly apply.  Instead, it is instead modeled as a system of equations that evaluates the temperature of all foam filters at once.  The radiative model is achieved entirely through balance of absorption and reflection, assuming no significant load transmitted through the filters.  Room temperature measurements have shown that the transmittance is very near 0 at the wavelengths dominating the thermal radiation, while the loads are relatively small at the low frequencies where the transmission is large for the polyethylene based HD-30 foam that we use, as it was for the polystyrene foam in the Choi. et al. paper (Fig. \ref{fig:hd30fts}) \cite{Wandui_thesis, Choi_2013}.

The results of this model depend strongly on the parameters for transmittance and conductance that are put into the model for each of the materials.  The model has references for alumina \cite{Inoue_2014}, HDPE \cite{Goldsmith}, fused quartz \cite{Inoue_2014}, teflon \cite{Goldsmith, NIST}, nylon \cite{NIST}, and Silicon \cite{Touloukian_1970, Afsar_1990}.  Where possible, cryogenic values have been used; however these parameters are not available for all materials at all temperatures.  Optical testing efforts at the Harvard labs aim to provide cryogenic references for more of these components, so that the reliability of this model may be improved.

   \begin{figure} [t]
   \begin{center}
   \begin{tabular}{cc}
   \includegraphics[height = 3cm]{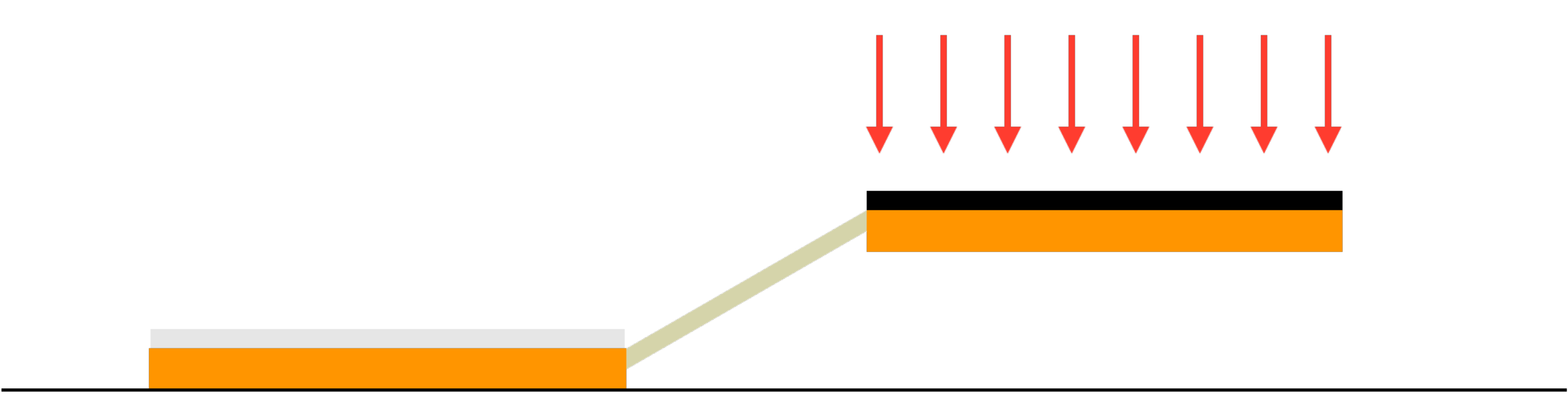} \\
   \includegraphics[height = 5cm]{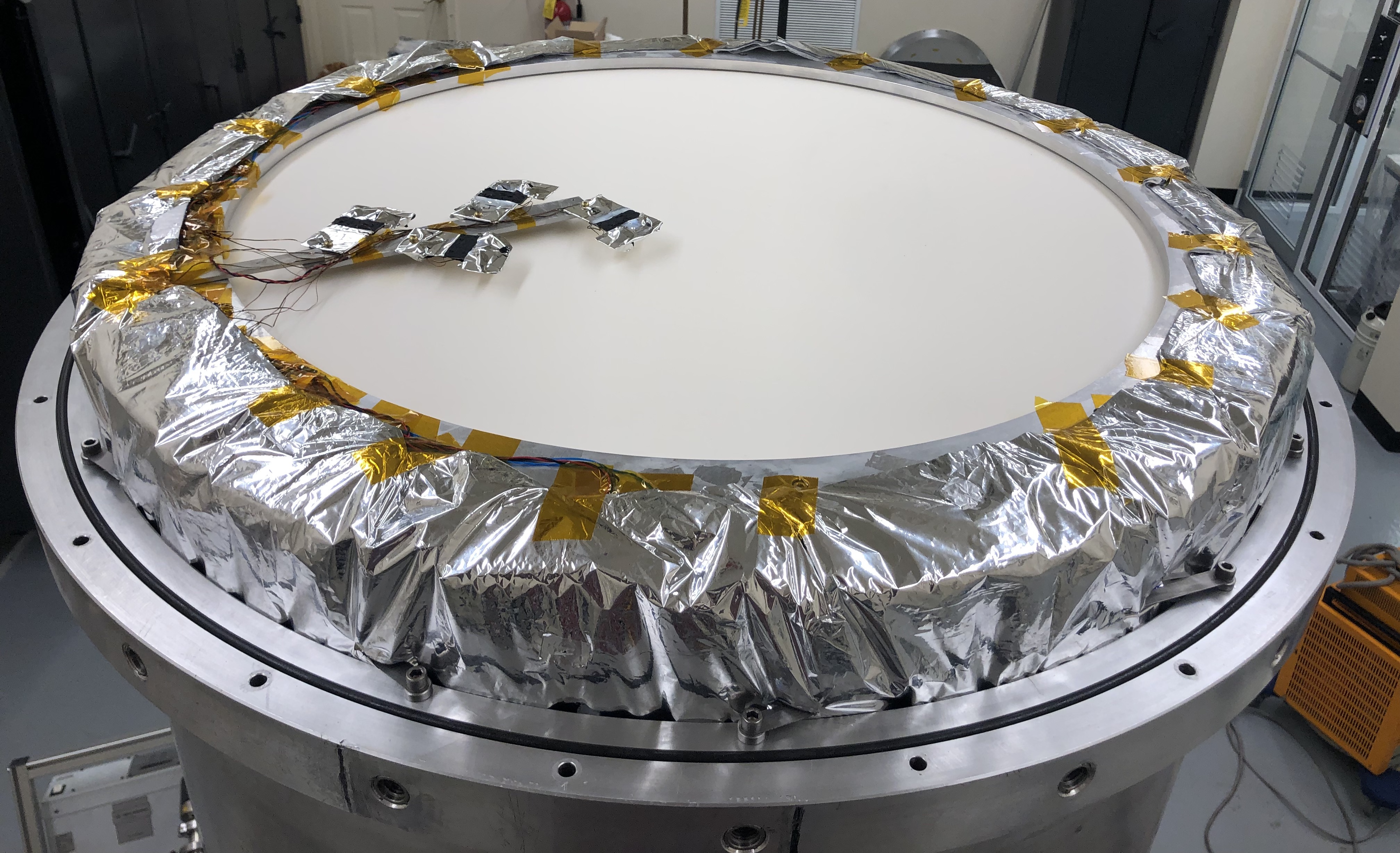}
   \includegraphics[height = 5cm]{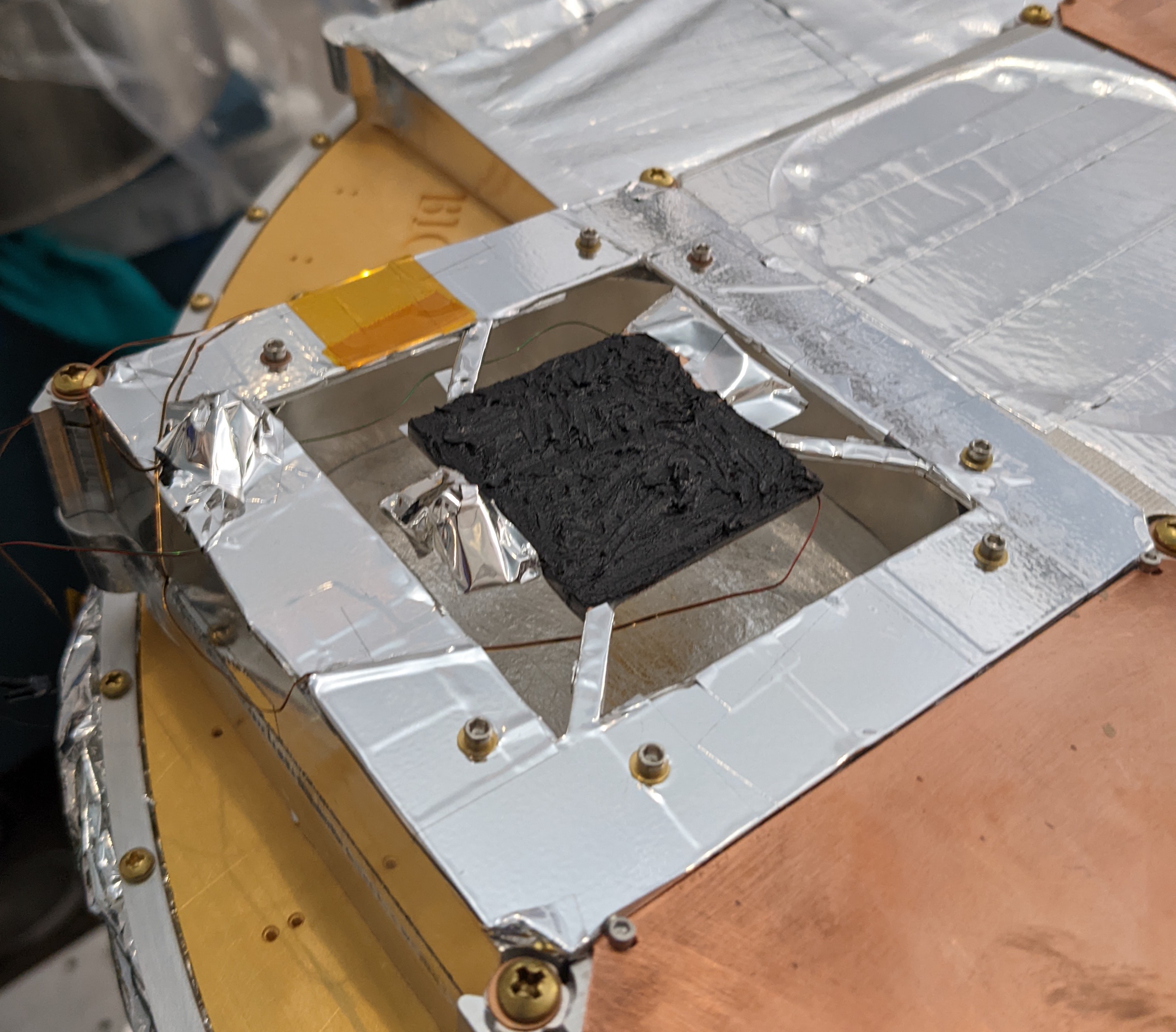} \\
   \end{tabular}
   \end{center}
   \caption[example] 
   { \label{fig:macrobolo} 
(Top) Diagram representing the macrobolometer layout, with the high emissivity absorber collecting incoming radiation, supported by thermally resistive material to provide a temperature difference. (Bottom left) Image of two macrobolometers, as installed in a BICEP Array cryostat above the 50K alumina filter.  (Bottom right) Image of one macrobolometer, as installed in a BICEP Array cryostat on the focal plane.}
   \end{figure}

\section{Thermal Testing}
\begin{figure} [t]
   \begin{center}
   \begin{tabular}{c}
   \includegraphics[height=4cm]{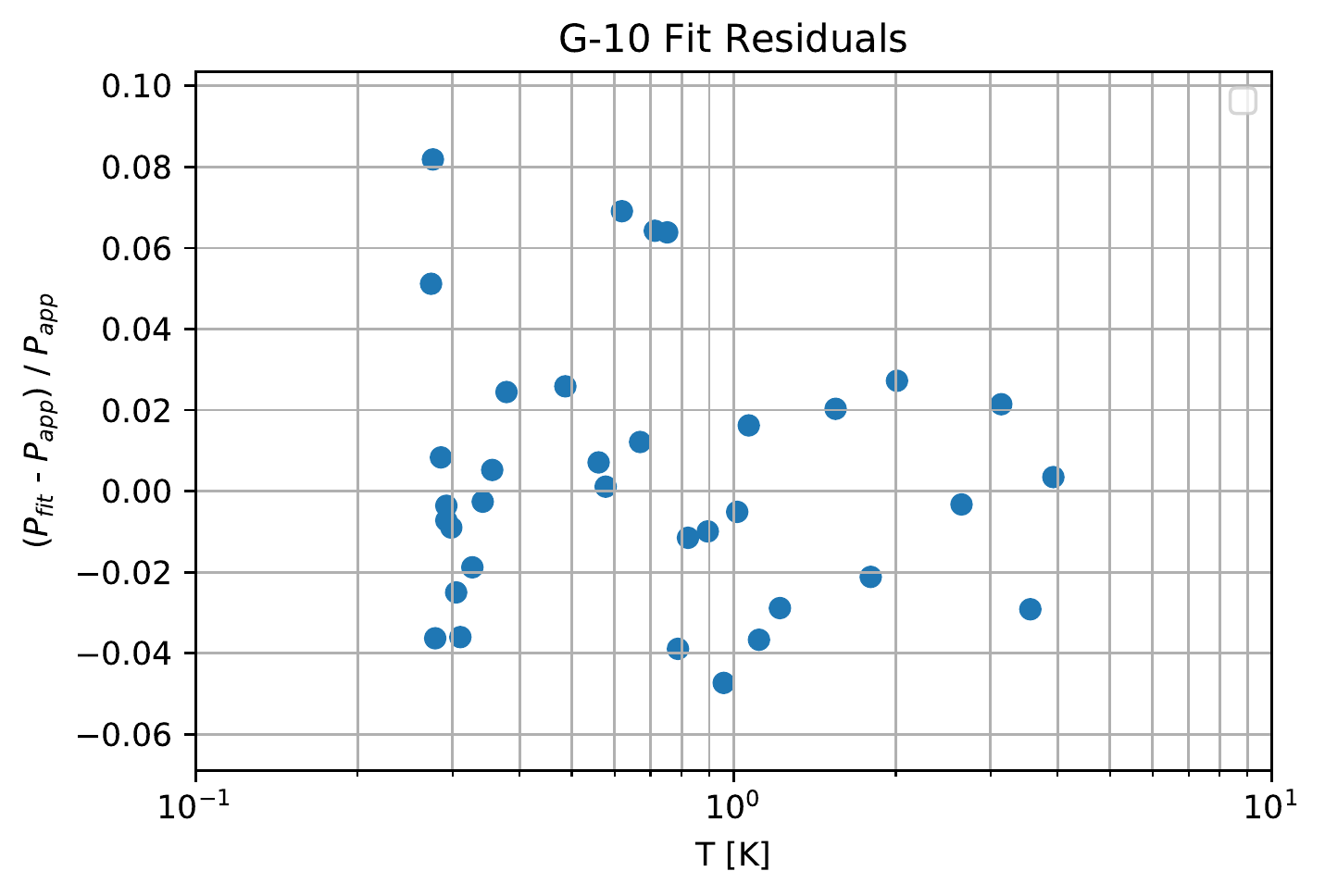}
   \end{tabular}
   \end{center}
   \caption[example] 
   { \label{fig:g10fit} 
Residuals comparing new measurements of the conductance of G10/FR4 samples with the fit to the conductance generated in Runyan and Jones\cite{Runyan_2008}.  These residuals are within the 10\% precision quoted in that paper, extending below the initial fit range to 0.27 K.}
   \end{figure}

In order to test how the performance of as-built cryogenic receivers compares to the thermal model, we have implemented cryogenic macrobolometers (Fig. \ref{fig:macrobolo}) that can be installed on different stages of the receiver to measure the incoming radiative load.  These macrobolometers work by taking an absorber varying in size from 2-10 cm$^2$ and supporting it with a section of thermally resistive G10/FR4 so that the absorbed thermal load will produce a measurable temperature difference between the two sides of the thermal resistance.  To ensure that the absorbing region has an emissivity of 1, the absorbing surface is covered with Bock Black \cite{BockThesis} while the rest of the device is wrapped in a low-emissivity aluminized mylar tape to minimize other hidden contributions to the absorber area.  The conductance of the thermally resistive portion is  measured in situ by applying additional heat to the absorber side of the device and measuring how the temperature difference scales with the load.  While fitting to that data, we assume a functional form for the temperature dependence of the conductivity that matches standard references \cite{NIST, Runyan_2008}.  Cryogenic measurements of our supply of G10/FR4 material without the macrobolometer absorbers has been shown to effectively match the reference in the 0.27 - 4.2 K range (Fig. \ref{fig:g10fit}).

\begin{figure} [ht]
   \begin{center}
   \begin{tabular}{cc}
   \includegraphics[height = 5cm]{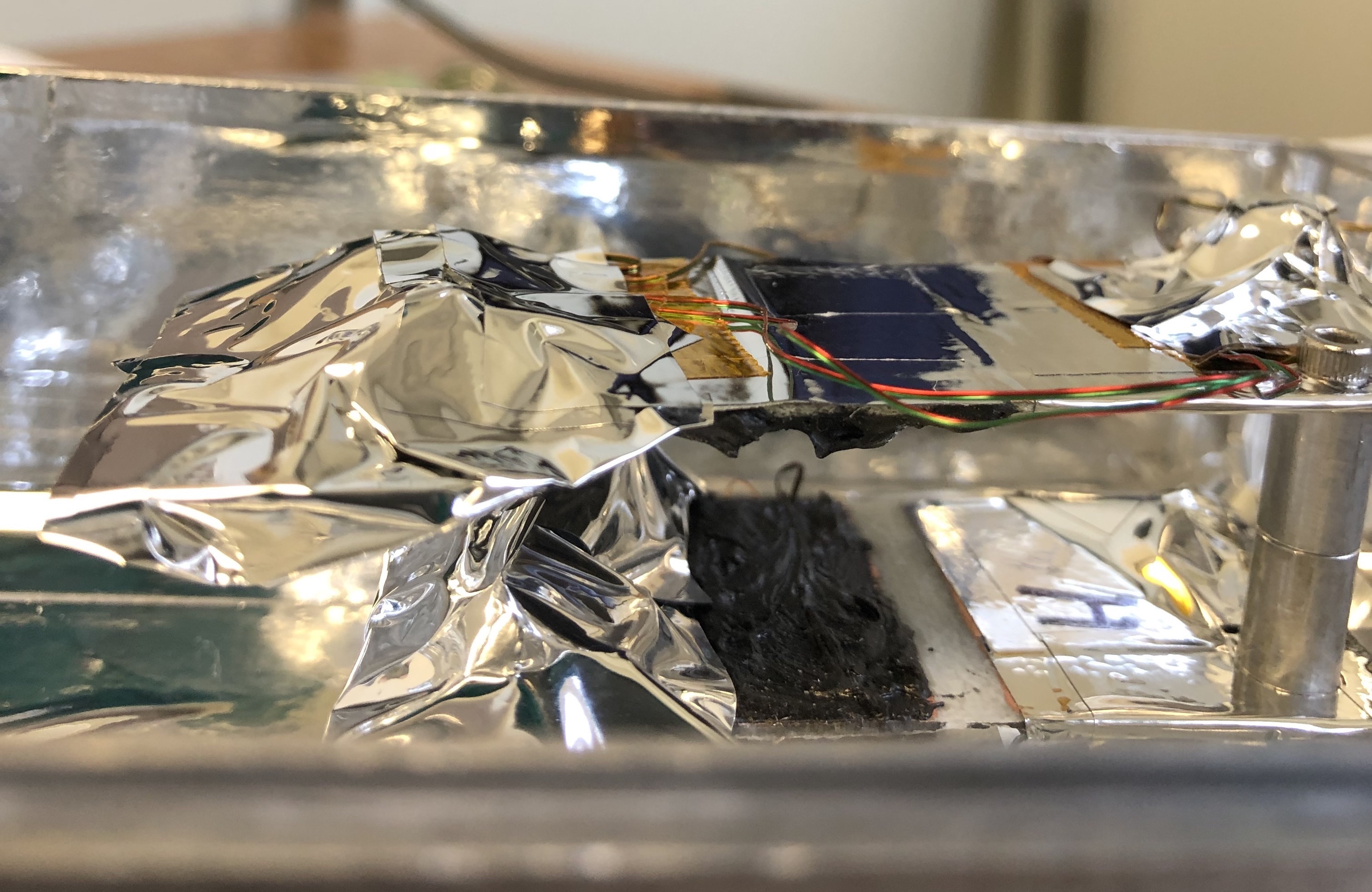}
   \includegraphics[height = 5cm]{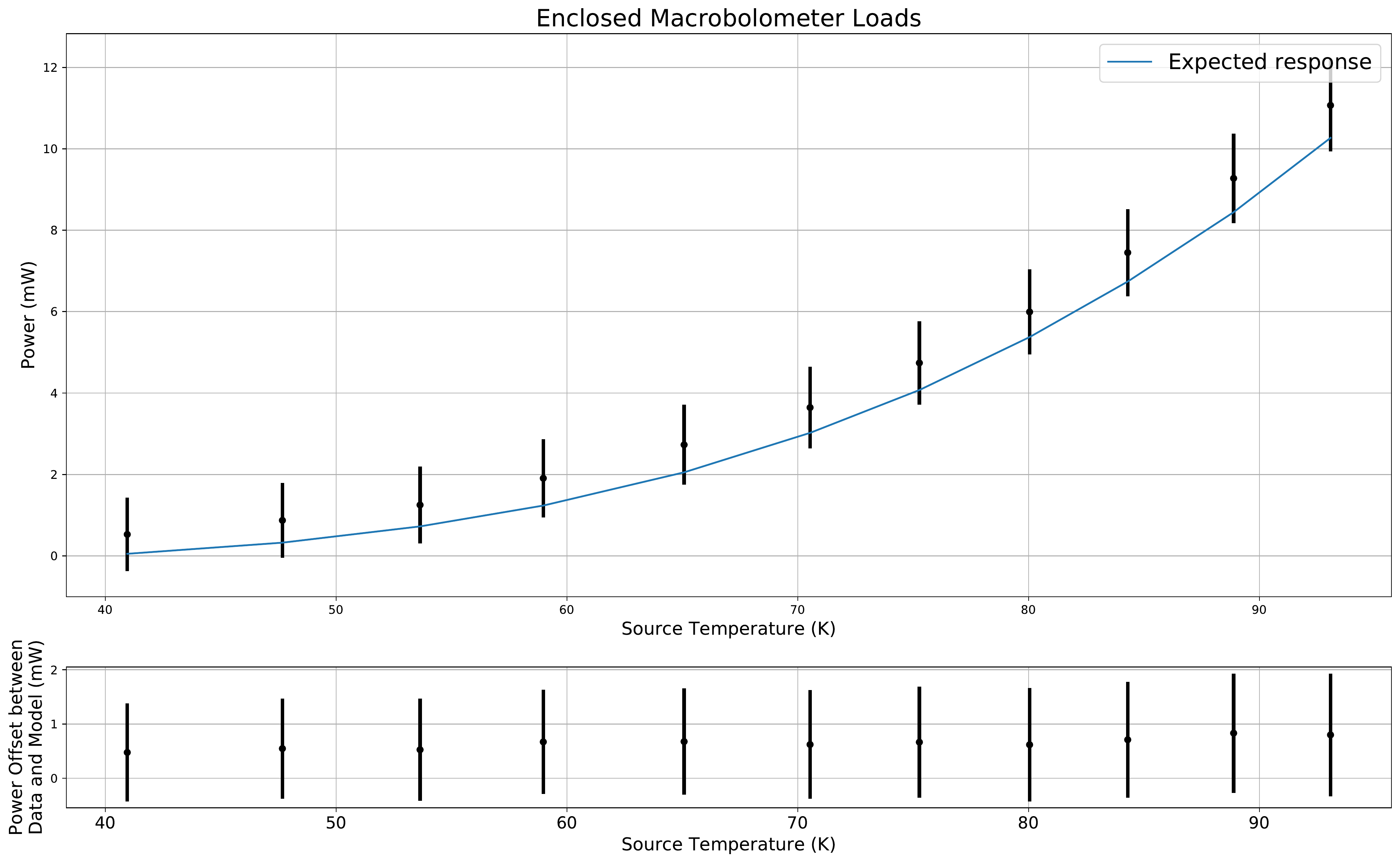}
   \end{tabular}
   \end{center}
   \caption[example] 
   { \label{fig:enclosure} 
(Left) Photograph of a miniature version of this setup for easier visualization. In the actual test, the devices are much larger (filling most of the enclosure area) and the devices are much closer together (as close as possible without risking contact) so that the source filled a larger solid angle.  (Right) Radiative load measured by the receiver macrobolometer, compared to the temperature of the source macrobolometer. The expected curve is calculated from the device geometries.}
   \end{figure}

To confirm that these macrobolometers can deliver physically reasonable measurements, we have installed two of these in an enclosure with the absorbing areas facing one another, such that the load incident on one device is dominated by the load radiated by the other, which could be controlled via a heater (Fig. \ref{fig:enclosure} Left).  Since we are measuring the temperature of the radiating region of the source macrobolometer, this allows us to provide a known load, scaled by the view factors between the two devices.  The results of this test are as expected, with the predicted load falling within the measurement range, taking into account the uncertainties arising from the precision of the thermometer calibration (Fig. \ref{fig:enclosure} Right).

\begin{figure} [ht]
   \begin{center}
   \begin{tabular}{c}
   \includegraphics[height=7cm]{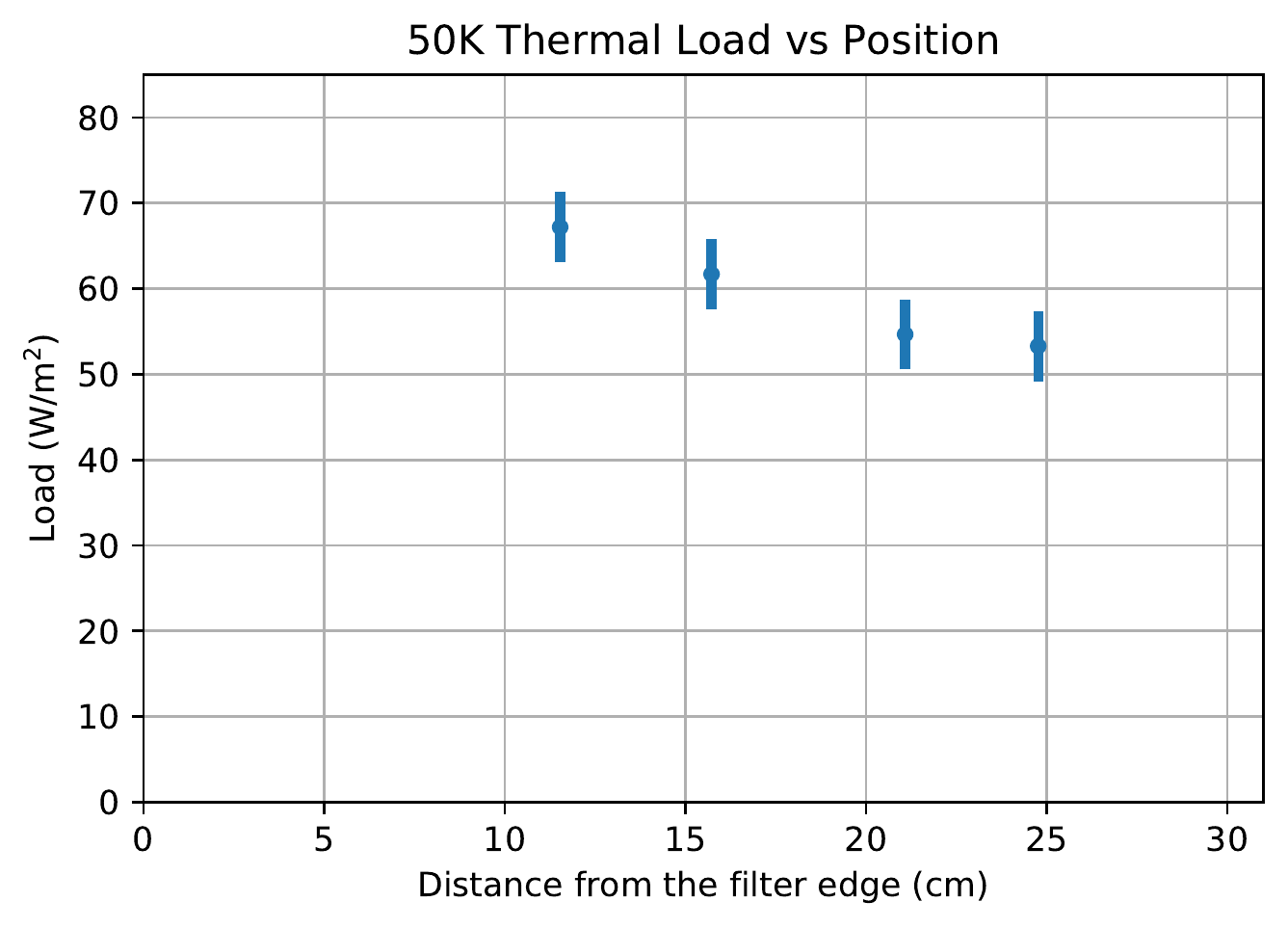}
   \end{tabular}
   \end{center}
   \caption[example] 
   { \label{fig:50kload} 
Radiative load incident on the 50K alumina filter, measured at different radial positions.}
   \end{figure}

These macrobolometers have been deployed in the BA3 receiver as it is being commissioned for its eventual deployment as part of BICEP Array.  By installing several of these above the 50K alumina filter, we are able to evaluate whether the performance of the foam filter stack is consistent with our expectations.  In fact, what we have measured is that the radiative load is significantly higher than expected, and is  higher near the edge of the filter than it is closer to the center (Fig. \ref{fig:50kload}).  The thermal model asserts that there will be a uniform load from the foam filter stack, based on the assumption that there is negligible conducted heat from the supports to the filter, so each surface should be isothermal.  This radial trend suggests that the support rings at the edge of the foam filters are conducting measurable heat from the vacuum jacket at room temperature, to provide the unmodeled load.  To test this conclusion, we have modified the support structure with the goal of mitigating this effect and reducing the overall load on our 50K stage.

Similar macrobolometers have been installed on the focal plane to measure the load coming from the 4K optics.  These devices have a larger absorbing area and a narrower resistive region than the 50K devices, in order to be sensitive to the smaller flux at that stage.  Using the BA2 receiver, which is equipped with the 4K optics, we have measured a load of $10.2\pm2.8$ $\mu$W/m$^2$.  The predicted load from the thermal model for this cryostat configuration is 11.2 $\mu$W/m$^2$, within the uncertainties of our measurement and providing confidence in our model at that level.

\section{Conclusions}
Astronomical measurements that requires cryogenic instrumentation to achieve the necessary sensitivity requires careful consideration of the thermal design to ensure that the cryostat achieves the required thermal performance.  We have developed a model of thermal radiation that passes through the optics of our receivers, which has been used to evaluate the performance of the BICEP Array telescope.  To verify the results of this model, we have also deployed macrobolometers to directly measure the radiation environment inside the cryostat.  This has allowed us to measure effects that are inconsistent with the assumptions of the model, motivating practical modifications to the design that can lead to future improvements in the performance.

\section{Acknowledgments}
The BICEP/Keck project (including BICEP2, BICEP3, and BICEP Array) have been made possible through a series of grants from the National Science Foundation including 0742818, 0742592, 1044978, 1110087, 1145172, 1145143, 1145248, 1639040, 1638957, 1638978, 1638970, 1726917, 1313010, 1313062, 1313158, 1313287, 0960243, 1836010, 1056465, \& 1255358 and by the Keck Foundation. The development of antenna-coupled detector technology was supported by the JPL Research and Technology Development Fund and NASA Grants 06-ARPA206- 0040, 10-SAT10-0017, 12-SAT12-0031, 14-SAT14-0009, 16-SAT16-0002, \& 18-SAT18-0017. The development and testing of focal planes were supported by the Gordon and Betty Moore Foundation at Caltech. Readout electronics were supported by a Canada Foundation for Innovation grant to UBC. The computations in this paper were run on the Odyssey cluster supported by the FAS Science Division Research Computing Group at Harvard University. The analysis effort at Stanford and SLAC was partially supported by the Department of Energy, Contract DE-AC02-76SF00515. We thank the staff of the U.S. Antarctic Program and in particular the South Pole Station without whose help this research would not have been possible. Tireless administrative support was provided by Kathy Deniston, Sheri Stoll, Irene Coyle, Amy Dierker, Donna Hernandez, and Julie Shih.

\bibliography{goldfinger_spie_2022.bib}

\begin{thebibliography}{10}

\bibitem{Crumrine_2018}
{Crumrine}, M., {Ade}, P.~A.~R., {Ahmed}, Z., {Aikin}, R.~W., {Alexander},
  K.~D., {Barkats}, D., {Benton}, S.~J., {Bischoff}, C.~A., {Bock}, J.~J.,
  {Bowens-Rubin}, R., {Brevik}, J.~A., {Buder}, I., {Bullock}, E., {Buza}, V.,
  {Connors}, J., {Cornelison}, J., {Crill}, B.~P., {Dierickx}, M., {Duband},
  L., {Dvorkin}, C., {Filippini}, J.~P., {Fliescher}, S., {Grayson}, J.~A.,
  {Hall}, G., {Halpern}, M., {Harrison}, S.~A., {Hildebrandt}, S.~R., {Hilton},
  G.~C., {Hui}, H., {Irwin}, K.~D., {Kang}, J.~H., {Karkare}, K.~S., {Karpel},
  E., {Kaufman}, J.~P., {Keating}, B.~G., {Kefeli}, S., {Kernasovskiy}, S.~A.,
  {Kovac}, J.~M., {Kuo}, C.~L., {Larsen}, N.~A., {Lau}, K., {Leitch}, E.~M.,
  {Lueker}, M.~V., {Megerian}, K.~G., {Moncelsi}, L., {Namikawa}, T.,
  {Netterfield}, C.~B., {Nguyen}, H.~T., {O'Brient}, R., {Ogburn}, R.~W.,
  {Palladino}, S., {Pryke}, C., {Racine}, B., {Richter}, S., {Schwarz}, R.,
  {Schillaci}, A., {Sheehy}, C.~D., {Soliman}, A., {St. Germaine}, T.,
  {Staniszewski}, Z.~K., {Steinbach}, B., {Sudiwala}, R.~V., {Teply}, G.~P.,
  {Thompson}, K.~L., {Tolan}, J.~E., {Tucker}, C.~E., {Turner}, A.~D.,
  {Umilt{\`a}}, C., {Vieregg}, A.~G., {Wandui}, A., {Weber}, A.~C., {Wiebe},
  D.~V., {Willmert}, J., {Wong}, C.~L., {Wu}, W.~L.~K., {Yang}, E., {Yoon},
  K.~W., and {Zhang}, C., ``{BICEP Array cryostat and mount design},'' in [{\em
  Millimeter, Submillimeter, and Far-Infrared Detectors and Instrumentation for
  Astronomy IX}{\nolinebreak\hspace{0.1em}]},  {Zmuidzinas}, J. and {Gao},
  J.-R., eds., {\em Society of Photo-Optical Instrumentation Engineers (SPIE)
  Conference Series} {\bf 10708},  107082D (July 2018).

\bibitem{Moncelsi_2020}
{Moncelsi}, L., {Ade}, P.~A.~R., {Ahmed}, Z., {Amiri}, M., {Barkats}, D., {Basu
  Thakur}, R., {Bischoff}, C.~A., {Bock}, J.~J., {Buza}, V., {Cheshire}, J.~R.,
  {Connors}, J., {Cornelison}, J., {Crumrine}, M., {Cukierman}, A.~J.,
  {Denison}, E.~V., {Dierickx}, M., {Duband}, L., {Eiben}, M., {Fatigoni}, S.,
  {Filippini}, J.~P., {Goeckner-Wald}, N., {Goldfinger}, D., {Grayson}, J.~A.,
  {Grimes}, P., {Hall}, G., {Halpern}, M., {Harrison}, S.~A., {Henderson}, S.,
  {Hildebrandt}, S.~R., {Hilton}, G.~C., {Hubmayr}, J., {Hui}, H., {Irwin},
  K.~D., {Kang}, J.~H., {Karkare}, K.~S., {Kefeli}, S., {Kovac}, J.~M., {Kuo},
  C.~L., {Lau}, K., {Leitch}, E.~M., {Megerian}, K.~G., {Minutolo}, L.,
  {Nakato}, Y., {Namikawa}, T., {Nguyen}, H.~T., {O'brient}, R., {Palladino},
  S., {Precup}, N., {Prouve}, T., {Pryke}, C., {Racine}, B., {Reintsema},
  C.~D., {Schillaci}, A., {Schmitt}, B.~L., {Soliman}, A., {St. Germaine}, T.,
  {Steinbach}, B., {Sudiwala}, R.~V., {Thompson}, K.~L., {Tucker}, C.,
  {Turner}, A.~D., {Umilt{\`a}}, C., {Vieregg}, A.~G., {Wandui}, A., {Weber},
  A.~C., {Wiebe}, D.~V., {Willmert}, J., {Wu}, W.~L.~K., {Yang}, E., {Yoon},
  K.~W., {Young}, E., {Yu}, C., {Zeng}, L., {Zhang}, C., and {Zhang}, S.,
  ``{Receiver development for BICEP Array, a next-generation CMB polarimeter at
  the South Pole},'' in [{\em Society of Photo-Optical Instrumentation
  Engineers (SPIE) Conference Series}{\nolinebreak\hspace{0.1em}]},  {\em
  Society of Photo-Optical Instrumentation Engineers (SPIE) Conference Series}
  {\bf 11453},  1145314 (Dec. 2020).

\bibitem{Crumrine_thesis}
Crumrine, M., {\em BICEP Array : Searching for Signals of Inflation from the
  South Pole}, PhD thesis, University of Minnesota (2022).

\bibitem{Choi_2013}
Choi, J., Ishitsuka, H., Mima, S., Oguri, S., Takahashi, K., and Tajima, O.,
  ``Radio-transparent multi-layer insulation for radiowave receivers,'' {\em
  Review of Scientific Instruments}~{\bf 84},  114502 (nov 2013).

\bibitem{Wandui_thesis}
Wandui, A., {\em Characterizing Infrared Filters for Large Aperture Millimeter
  Wave Telescopes}, bachelor's thesis, Stanford University (2017).

\bibitem{Inoue_2014}
Inoue, Y., Matsumura, T., Hazumi, M., Lee, A.~T., Okamura, T., Suzuki, A.,
  Tomaru, T., and Yamaguchi, H., ``Cryogenic infrared filter made of alumina
  for use at millimeter wavelength,'' {\em Applied Optics}~{\bf 53},  1727 (mar
  2014).

\bibitem{Goldsmith}
Goldsmith, P.,  [{\em Quasioptical Systems}{\nolinebreak\hspace{0.1em}]}, IEEE
  Press, New York (1998).

\bibitem{NIST}
{NIST}, ``Index of material properties.''
  \url{https://trc.nist.gov/cryogenics/materials/materialproperties.htm}.

\bibitem{Touloukian_1970}
Touloukian, Y.~S., Powell, R., Ho, C., and Klemens, P.~G., ``Thermal
  conductivity of metallic elements and alloys,'' {\em Thermophysical
  Properties of Matter}~{\bf 1},  333--339 (1970).

\bibitem{Afsar_1990}
Afsar, M., Chi, H., and Li, X., ``Millimeter wave complex refractive index,
  complex dielectric permittivity and loss tangent of high purity and
  compensated silicon,''  238--239 (1990).

\bibitem{BockThesis}
Bock, J., PhD thesis, University of California at Berkeley (1994).

\bibitem{Runyan_2008}
Runyan, M. and Jones, W., ``Thermal conductivity of thermally-isolating
  polymeric and composite structural support materials between 0.3 and 4k,''
  {\em Cryogenics}~{\bf 48},  448--454 (sep 2008).

\end{thebibliography}
\bibliographystyle{spiebib} 

\end{document}